\documentclass[aps,prl,twocolumn,showpacs,superscriptaddress]{revtex4-1}  
\usepackage{float}
\usepackage{bbm}
\usepackage{epsfig}
\usepackage{epstopdf}
\usepackage{graphicx}
\usepackage{amsmath,amssymb}
\usepackage{amsmath,bm}
\usepackage{physics}
\usepackage{color}
\usepackage{hyperref}
\usepackage{lineno,blindtext}
\usepackage[caption=false]{subfig}
\usepackage{siunitx, booktabs}
\usepackage{diagbox, eqparbox, hhline}
\usepackage{soul}
\usepackage{xcolor}
\usepackage[normalem]{ulem}
\newcolumntype{P}[1]{>{\centering\arraybackslash}p{#1}}

\begin{document}

\title{{Exceptional points as signatures of dynamical magnetic phase transitions}}

\author{Kuangyin Deng}
\email{dengku@bc.edu}
\author{Xin Li}
\author{Benedetta Flebus}
\email{flebus@bc.edu}
\affiliation{Department of Physics, Boston College, 140 Commonwealth Avenue, Chestnut Hill, Massachusetts 02467, USA}

\begin{abstract}

One of the most fascinating and puzzling aspects of non-Hermitian systems is their spectral degeneracies, i.e., exceptional points (EPs), at which both eigenvalues and eigenvectors coalesce to form a defective state space. While coupled magnetic systems are natural hosts of EPs, the relation between the linear and nonlinear spin dynamics in the proximity of EPs remains relatively unexplored. Here we theoretically investigate the spin dynamics of easy-plane magnetic bilayers in the proximity of exceptional points. We show that the interplay between the intrinsically dissipative spin dynamics and external drives can yield a rich dynamical phase diagram. In particular, we find that, in antiferromagnetically coupled bilayers, a periodic oscillating dynamical phase emerges in the region enclosed by EPs. Our results not only offer a pathway for probing magnetic EPs  and engineering magnetic nano-oscillators with  large-amplitude oscillations, but also uncover the relation between exceptional points and dynamical phase transitions in systems displaying non-linearities.

\end{abstract}

\maketitle


\textit{Introduction}. The degeneracies of Hermitian Hamiltonians are diabolic points, i.e.,  points at which two (or more) real eigenenergies coalesce, while the eigenstates still span the full Hilbert space. 
Non-Hermitian degeneracies, i.e., exceptional points (EPs),  display  properties that are radically different from their Hermitian counterpart. At an EP,  two (or more) complex eigenvalues and the corresponding eigenvectors simultaneously coalesce, resulting into a defective Hamiltonian that cannot span the entire Hilbert space~\cite{heiss2000repulsion,heiss2012physics,dembowski2001experimental}. 
The incompleteness of the eigenbases at second-order EPs leads to  a square root dependence on external perturbations, resulting in a giant sensitivity-factor enhancement ~\cite{chen2017exceptional,wiersig2014enhancing,miri2019exceptional,park2020symmetry}.

As non-Hermitian systems are recently under comprehensive research~\cite{gong2018topological,shen2018topological,yao2018edge,kawabata2019symmetry,bergholtz2021exceptional}, intense efforts have been put forward to  explore the properties of EPs. Particular emphasis has been placed on $\mathcal{PT}$-symmetric systems~\cite{ozdemir2019parity,xiao2021observation,miri2019exceptional,achilleos2017non},  where EPs signal a $\mathcal{PT}$-symmetry-breaking transition at which a system's eigenvalues turn from real to complex conjugate pairs. The emergence of EPs does not, however, require a fine-tuned balance of gain and loss~\cite{zhiyenbayev2019enhanced}. EPs have been reported in a plethora of open systems, ranging from optics and photonics~\cite{miri2019exceptional,chen2017exceptional,ozdemir2019parity,longhi2014optical,renault2019virtual} to superconducting quantum circuits~\cite{partanen2019exceptional}, semimetals~\cite{molina2018surface,kawabata2019classification,gonzalez2017topological,chowdhury2022exceptional}, and magnetic systems~\cite{komineas2022non,li2022multitude,lee2015macroscopic,galda2016parity,yang2018antiferromagnetism,yu2020higher,hurst2022perspective,wang2021enhanced,jeffrey2021effect,flebus2020non,Gunnink2022,liu2019observation}. 

Magnetic systems are intrinsically open   due to the ubiquitous dissipation of magnetization dynamics~\cite{tserkovnyak2020exceptional,liu2019observation,deng2022non}. 
The gain can be tuned via experimentally established techniques such as, e.g., spin current injection~\cite{flebus2021magnonics,liu2019observation,ralph2008spin,sun2008magnetoresistance,katine2008device,chen2016spin}. Exceptional points naturally emerge in the description of coupled magnetization dynamics and have been recently observed in magnonic $\mathcal{PT}$-symmetric devices~\cite{liu2019observation}. Second-order and higher-order EPs displaying  higher-order roots singularities~\cite{zhang2019higher,mandal2021symmetry,wang2019arbitrary,hodaei2017enhanced,wu2021high,xiao2019enhanced,zhang2019quantum},  which can yield further  ultra-sensitivity, have been reported in magnetic multilayers~\cite{yu2020higher}.
While the potential of EPs in magnetic sensing has been under intense scrutiny, the role that EPs play in dynamical magnetic phase transitions is yet relatively unexplored. 

Coupled magnetization dynamics can be described, in the long-wavelength limit, via the coupled Landau-Lifshitz-Gilbert (LLG) equations~\cite{lakshmanan2011fascinating}. By linearizing the LLG equations of motion, one can derive an effective non-Hermitian Hamiltonian quadratic in second-quantized magnon operators. The EPs appear as singularities of the quadratic Hamiltonian, signaling a dynamical phase transition of the linearized dynamics due to a width bifurcation~\cite{rotter2009non,eleuch2016clustering,eleuch2013width,rotter2010exceptional}. If signatures of such transition survive in the nonlinear LLG-like classical dynamics, the analysis of the corresponding quadratic magnon Hamiltonian can  unveil unforeseen dynamical regimes as function of experimentally tunable parameters.

In this work, we explore the connection between linear and nonlinear spin dynamics in proximity of EPs by taking an easy-plane magnetic  bilayer as an example. The ratio between gain and loss is modulated by spin injection in the bottom layer and the loss of magnetization dynamics is taken to be larger than the overall gain.  As a function of the interlayer coupling, we find that the linearized spectrum displays two regions  encircled by exceptional points, emerging around, respectively, vanishing and strong antiferromagnetic (AFM) interlayer coupling. The non-linear dynamics in proximity of the region with vanishing interlayer coupling displays a ferromagnetic (FM)-to-AFM dynamical phase transition. Such transition has been reported in a magnonic $\mathcal{PT}$-symmetric system~\cite{yang2018antiferromagnetism}: our results show that fine-tuned balance of gain and loss is not necessary for the transition to take place.

Furthermore, we unveil a distinct dynamical phase transition occuring in the  AF-coupled region encircled by the EPs.  Simulations of the nonlinear  dynamics  show, that upon crossing the EP in parameter space, the damped magnetization dynamics enters a regime of steady self-oscillations with large amplitude that can be described by a supercritical Hopf-Bifurcation~\cite{kim2012spin,strogatz2018nonlinear,mayergoyz2009nonlinear}. According to our estimates,  this dynamical phase transition might be observed in van der Waals  and synthetic AFM bilayers  ~\cite{macneill2019gigahertz,sklenar2021self}, which could open up
a route to engineer magnetic nano-oscillators~\cite{chen2016spin,chen2020spin,markovic2019reservoir,houssameddine2007spin,kaka2005mutual,tiberkevich2007microwave,firastrau2013spin,cheng2016terahertz} with large-amplitude oscillations. Our findings have also the potential to shed light on the interplay between EPs and dynamical phase transition in other dissipative-driven systems displaying non-linearities.
\begin{figure}
\centering
{{\includegraphics[trim=0cm 0cm 0cm 0cm, clip=true,width=8.5cm, angle=0]{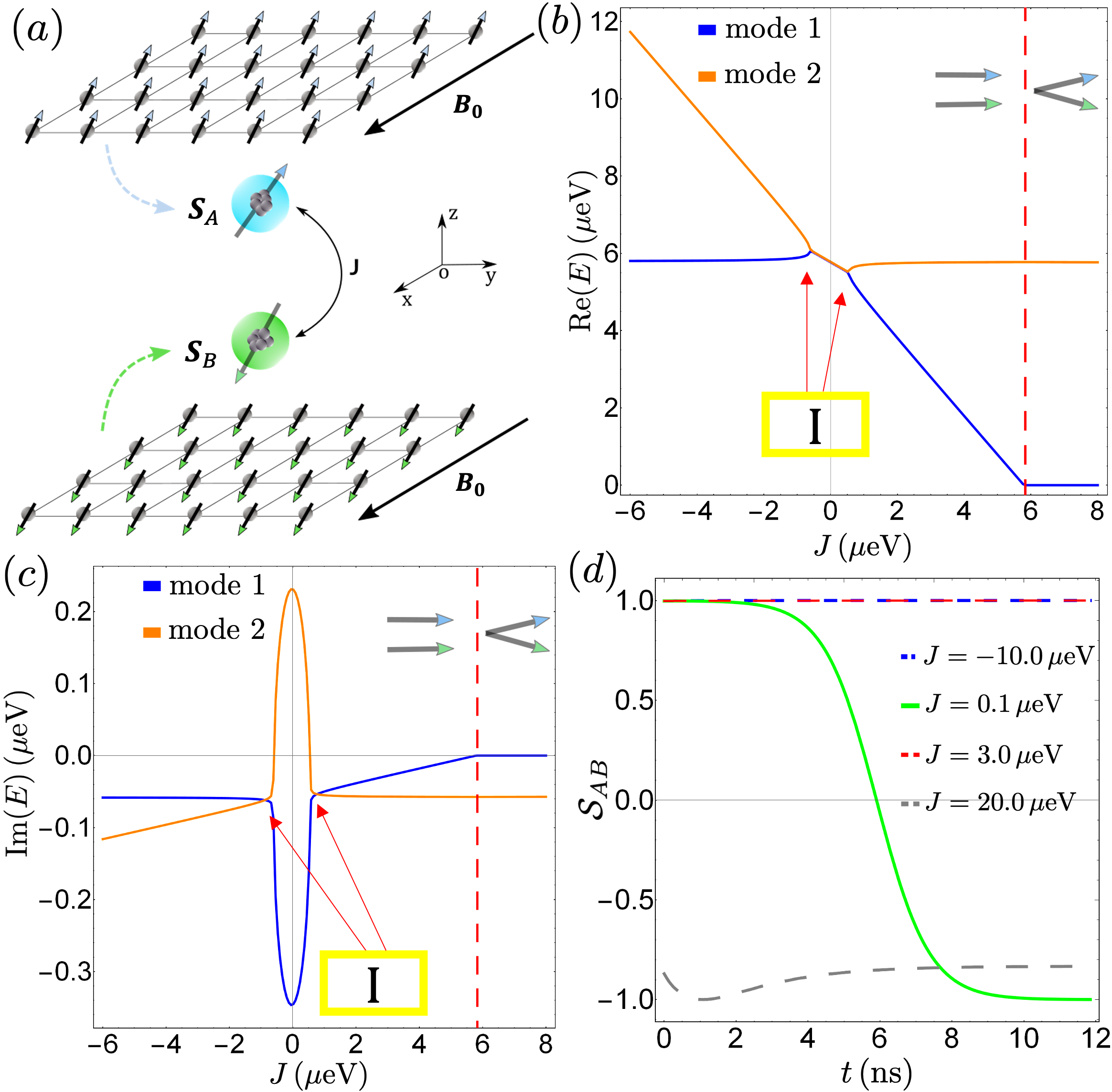}}}
\caption{(a) Magnetic bilayer with interlayer coupling $J$ in an external magnetic field $\mathbf{B}_{0}$. In the long-wavelength limit, the uniform magnetization of the top (bottom) layer can be treated as a macrospin $\mathbf{S}_{A(B)}$. (b,c): Dependence on $J$ of the real and imaginary energy, respectively,  for $K=0$. Region I  is enclosed by EPs. The red dashed line separates a collinear from a non-collinear ground state. (d) The time evolution of $\mathcal{S}_{AB}$  for different values of the interlayer coupling $J$. The FM-to-AFM dynamical phase transition emerges  in region I for small interlayer coupling, e.g., $J=0.1$ $\mu$eV. Instead, for values of $J$ further away from region I, the relative alignment of the macrospins remains the one of the corresponding ground state. In each figure, the parameters are set to $B_0=0.1$ T, $K=0$, $\alpha_A=0.06$ and $\alpha_B=-0.04$.}\label{Fig:1}
\end{figure}

\textit{Model}. We consider  the magnetic bilayer shown in Fig.~\hyperref[Fig:1]{\ref*{Fig:1}(a)}, whose  spin Hamiltonian can be written, in the long-wavelength limit,  as

\begin{align}
\mathcal{H}=&\sum_{i=A,B} \left(K S^{z\,2}_{i} + \gamma \mathbf{B}_{0} \cdot \mathbf{S}_{i}\right)+J\mathbf{S}_{A}\cdot\mathbf{S}_{B},\label{eq:simple-eg-1}
\end{align}
where  $\mathbf{S}_{A(B)}$, with $|\mathbf{S}_{A,B}|=S$, is the (dimensionless)  macrospin operator of the top (bottom) layer,
$\mathbf{B}_0$  the applied magnetic field, $\gamma>0$  the gyromagnetic, $J$ the interlayer coupling, and $K\ge0$  parametrizes the easy-plane anisotropy. Here we set $\hbar=1$ by adopting its unit to other parameters.  To introduce loss and gain, we recast the magnetization dynamics in the form of coupled Landau–Lifshitz–Gilbert (LLG) equations~\cite{lakshmanan2011fascinating}, i.e., 
\begin{align}
\frac{\mathrm{d}\mathbf{S}_A}{\mathrm{d}t} &=-\gamma\mathbf{S}_A\cross\mathbf{B}^{\text{eff}}_A-\frac{\alpha_A}{S}\mathbf{S}_A\cross\frac{\mathrm{d}\mathbf{S}_A}{\mathrm{d}t}\,, \label{eq:simple-eg-2a}\\
\frac{\mathrm{d}\mathbf{S}_B}{\mathrm{d}t} &=-\gamma\mathbf{S}_B\cross\mathbf{B}^{\text{eff}}_B-\frac{\alpha_B}{S}\mathbf{S}_B\cross\frac{\mathrm{d}\mathbf{S}_B}{\mathrm{d}t},\label{eq:simple-eg-2}
\end{align}
where we have introduced the effective field $\gamma\mathbf{B}^{\text{eff}}_{i}=\partial \mathcal{H}/\partial \mathbf{S}_{i}$, with $i=A,B$. Here $\alpha_{A}>0 $ $(\alpha_{B}<0)$ represents the effective damping (gain) parameter of the top (bottom) layer.  

To investigate the non-Hermitian spin-wave spectrum as function of the exchange coupling $J$ and magnetic field $\mathbf{B}_{0}$, we orient the spin-space Cartesian coordinate system such that the $\hat{\mathbf{z}}$ axis locally lies along the classical orientation of the macrospin $
\tilde{\mathbf{S}}_{i}$. The latter can be related to the spin operator $\mathbf{S}_{i}$ in the global frame of reference via the transformation~\footnote{See Supplementary Material for the coordinate transformation.}
\begin{align}
\mathbf{S}_{i}=\mathcal{R}_{z}(\phi_{i}) \mathcal{R}_{y}(\theta_{i}) \tilde{\mathbf{S}}_{i}\,,
\end{align}
where the matrix $\mathcal{R}_{z(y)}(\eta)$ describes a right-handed rotation by an angle $\eta$ about the $\hat{z} (\hat{y})$  axis, and $\theta_{i} (\phi_{i})$ is the polar (azimuthal) angle of the classical orientation of the spin $\mathbf{S}_{i}$. We then solve self-consistently Eqs.~(\ref{eq:simple-eg-2a}) and~(\ref{eq:simple-eg-2}) in the linear approximation, i.e., we consider $\tilde{\mathbf{S}}_{i}=\left( \tilde{S}_{i}^{x}, \tilde{S}_{i}^{y}, S \right)$.  Next, we introduce the complex variable $\tilde{S}^{+}_{i} = \tilde{S}^{x}_{i}+i \tilde{S}^{y}_{i}$ and invoke the Holstein-Primakoff transformation  $\tilde{S}^{+}_{A(B)}\approx\sqrt{2S} a(b)$, where the second-quantized operator $a(b)$ annihilates a magnon in the top (bottom) layer and obeys bosonic commutation relations~\cite{HPtrans}. By invoking the Heisenberg equation for $a(b)$, we obtain the non-Hermitian Hamiltonian $\mathcal{H}_{nh}$. The resulting Hamiltonian is not block-diagonal and a Bogoliubov transformation is required to obtained the spin-wave spectrum ~\footnote{See Supplementary Material for a detailed derivation.}

\textit{Antiferromagnetic to ferromagnetic transition.} As a first instructive example, we turn off the easy-plane anisotropy, i.e., $K=0$, and we take a damping coefficient of the same order of magnitude of the ones reported for chromium trihalide crystals~\cite{shen2021multi}, i.e.,  $\alpha_A=0.06$, while we set  $\alpha_B=-0.04$~\footnote{These values can be chosen with a certain flexibility as long as the loss is larger than the gain, as we  discuss in detail later.}. 
We set $B_0=0.1$ T and  take  $\mathbf{B}_0 \parallel \hat{x}$. It is worth noting that our results do not depend on the field direction since the Hamiltonian~\eqref{eq:simple-eg-1} is $SO(3)$-symmetric for $K=0$. The real and imaginary energy spectra of $\mathcal{H}_{nh}$ as a function of $J$ are shown, respectively, in Fig.~\hyperref[Fig:1]{\ref*{Fig:1}(b)} and~\hyperref[Fig:1]{\ref*{Fig:1}(c)}. Near $J=0$, region I is enclosed by EPs. On the left side of the red dashed line, the ground state of the Hermitian Hamiltonian  (i.e., Eq.~\eqref{eq:simple-eg-1} for $\alpha_{A(B)}=0$)  is collinear  and oriented along the magnetic field. On the right side of the dashed line, the interplay between the magnetic field and the  antiferromagnetic coupling $J$ leads to a noncollinear ground state, while increasing $J$ further yields an AFM ground state. 

To investigate how the degeneracies of the non-Hermitian linear spectrum affect the non-linear 
magnetization dynamics, we simulate  Eqs.~\eqref{eq:simple-eg-2a} and~\eqref{eq:simple-eg-2}  by setting the initial direction of the spins slightly away ($2^{\circ}$) from their  ground-state equilibrium position. We solve 
Eqs.~\eqref{eq:simple-eg-2a} and~\eqref{eq:simple-eg-2} for different values of $J
$ and track the time evolution of the product of the macrospins, i.e., $\mathcal{S}_{AB}(t)=\mathbf{S}_A(t) \cdot\mathbf{S}_B(t) /S^2$. As shown in Fig.~\hyperref[Fig:1]{\ref*{Fig:1}(d)},  the relative alignment $\mathcal{S}_{AB}$ between the macrospins remains FM or AFM for values of $J$ further away from the exceptional point. Instead when we chose $J$ within region I, we observe a switch from a FM to an AFM configuration. Our result agrees with the observations of Ref.~\cite{yang2018antiferromagnetism}, in which 
the authors analyze the $\mathcal{PT}$-symmetric case (i.e., $\alpha_{A}=-\alpha_{B}$) of Eqs.~(\ref{eq:simple-eg-1}-\ref{eq:simple-eg-2}) for $K=0$. Here, we propose a simple explanation for this dynamical phase transition, which occurs when the coupling $J$ is close to $0$. In this regime, the spins are barely coupled and, thus, eventually, each macrospin  obeys its individual dynamics.  The macrospin experiencing gain  flips, while the lossy one  recovers its equilibrium orientation, leading to an AFM  orientation. As we have shown, $\mathcal{PT}$ symmetry is not required for the FM-to-AFM switching to occur.

\begin{figure}
\centering
{{\includegraphics[trim=0cm 0cm 0cm 0cm, clip=true,width=8.6cm, angle=0]{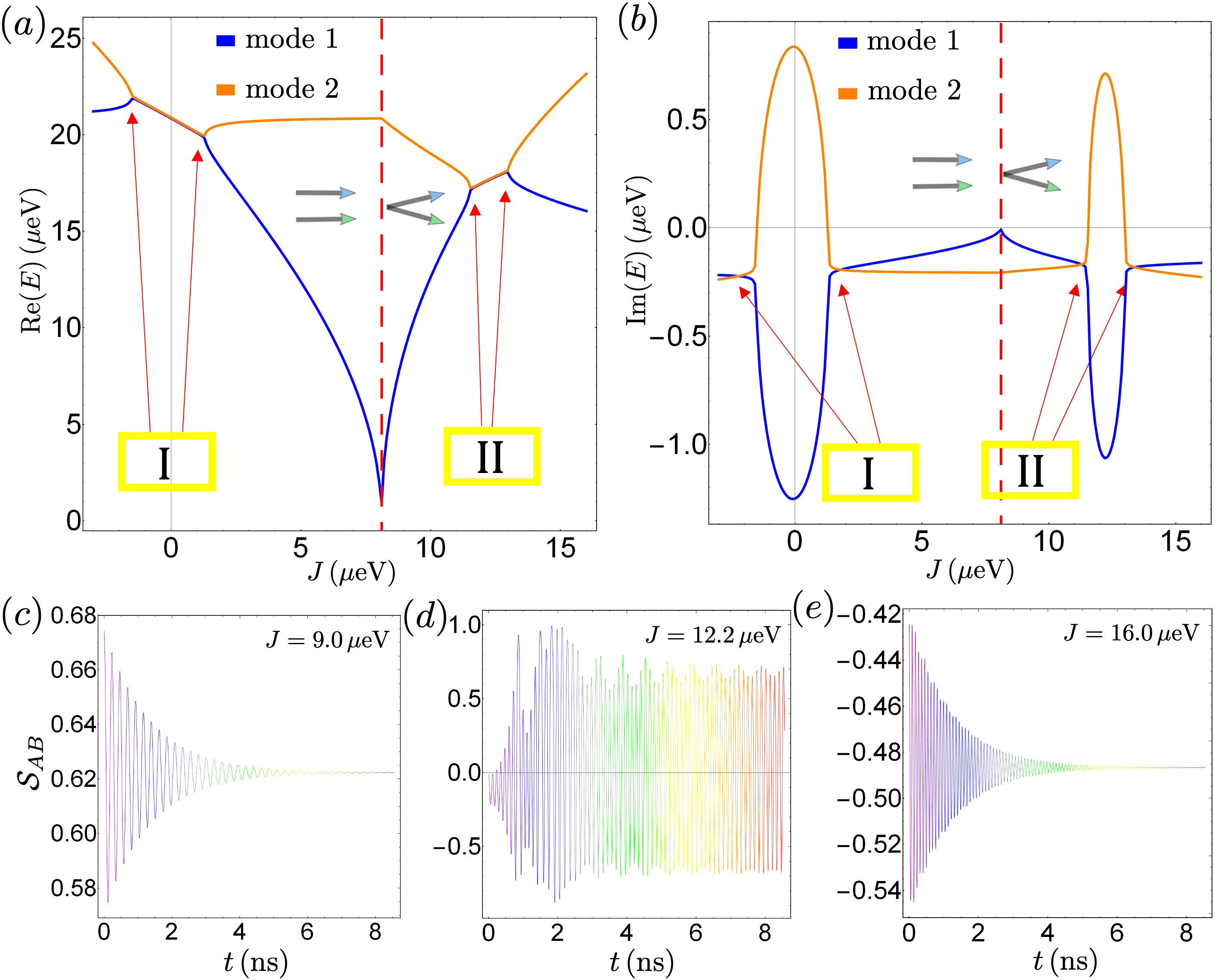}}}
\caption{Real (a) and imaginary (b)  energy for $B_0=0.14$ T, $K=45.9$ $\mu$eV, $\alpha_A=0.06$ and $\alpha_B=-0.04$. Here, region I is in direct correspondence with region I of Fig.~\hyperref[Fig:1]{\ref*{Fig:1}}. The red dashed line marks the transition from a collinear to a noncollinear configuration. Region II is enclosed by another pair of EPs in the noncollinear configuration. (c) - (e) The time evolution of $\mathcal{S}_{AB}(t)$ for different values of the interlayer coupling $J$. A periodic dynamical phase emerges only within region II.}\label{Fig:2}
\end{figure}

\begin{figure*}
\centering
{{\includegraphics[trim=1.0cm 0cm 0cm 0cm, clip=true,width=17.5cm, angle=0]{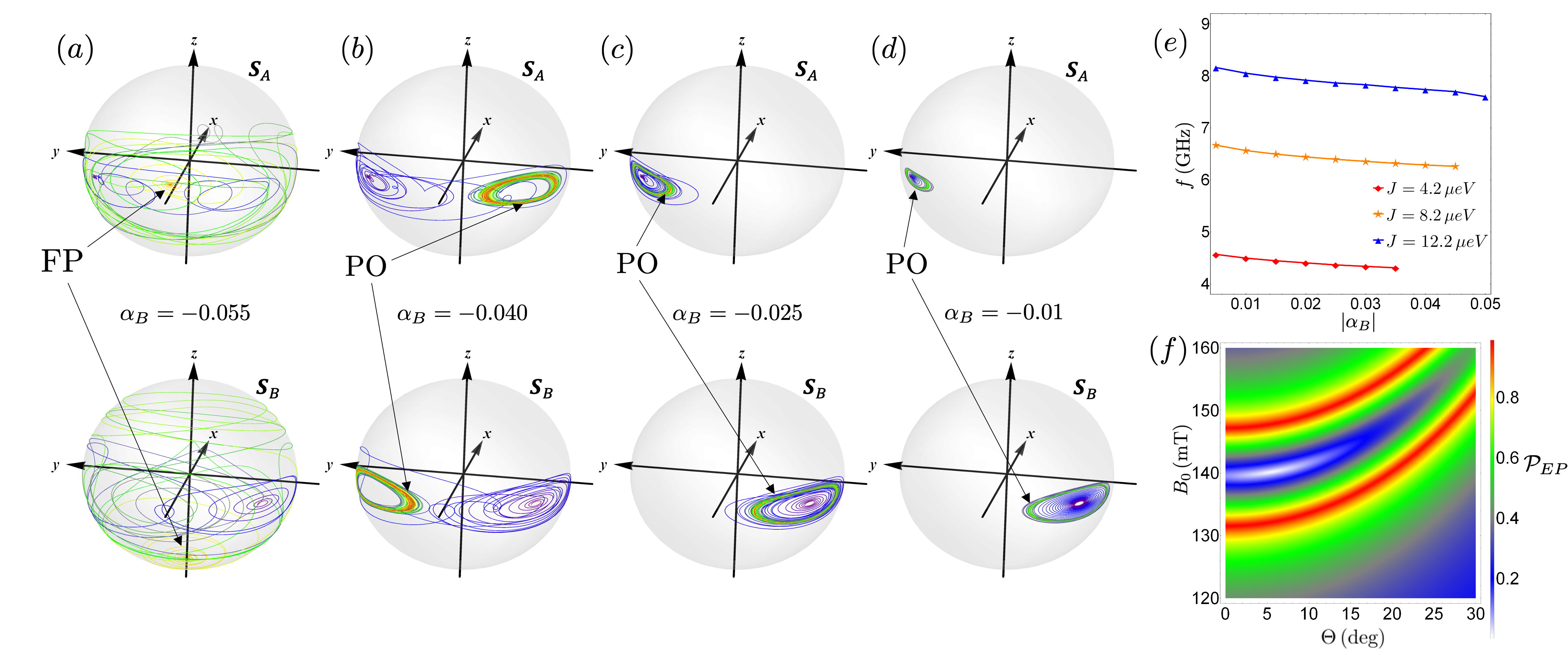}}}
\caption{(a) - (d) The spin evolution on the Bloch spheres for different values of the effective gain $\alpha_B$ in the region II of Fig.~\hyperref[Fig:2]{\ref*{Fig:2}} for $B_0=0.14$ T, $K=45.9$ $\mu$eV, $J=12.2$ $\mu$eV, and $\alpha_A=0.06$. The above (below) panels shows the time evolution of $\mathbf{S}_A$ ($\mathbf{S}_B$). The color on curves are in direct correspondence with the time intervals of the time-evolution of $\mathcal{S}_{AB}$ in Figs.2(c)-(d), i.e., they   label the earliest to the latest time  by ordering purple, blue, gray, green, yellow, orange, and red. (a) For $\alpha_{B}=-0.055$, the dynamics of $\mathbf{S}_{A}$ and $\mathbf{S}_{B}$ flow into fixed points (FP). (b-d): When $\abs{\alpha_B}\leq 0.05$, the system drops on periodic orbitals (PO) through the supercritical Hopf-Bifurcation. $\mathbf{S}_A$ with larger loss than the gain in $\mathbf{S}_B$ would form smaller orbitals to maintain the steady periodic oscillation. (e) Frequency $f$ of the coupled oscillations $\mathcal{S}_{AB}$ as a function of the effective gain $\alpha_B$ for different values of $J$. For $J=4.2$ $\mu$eV ($J=8.2$ $\mu$eV), steady periodic dynamical phases exist only for $\abs{\alpha_B}\leq 0.035$ ($\abs{\alpha_B}\leq 0.045$). (f) The dependence of the square of the overlap of the two right eigenvectors, i.e.,  $\mathcal{P}_{EP}\equiv \abs{\bra{\psi_1^R}\ket{\psi_2^R}}^2$, on the magnetic field strength $B_{0}$ and polar angle $\Theta$. } \label{Fig:3}
\end{figure*}

\textit{A magnetic nano-oscillator}. To explore the dynamical phase diagram of our model, we now turn on the easy-plane anisotropy, i.e., $K>0$. With  CrCl$_3$ in mind, we set   $K=45.9$ $\mu$eV~\cite{macneill2019gigahertz}. 
We consider a $U$(1)-symmetry breaking magnetic field $\mathbf{B}_0 \parallel \hat{x}$ and set $B_0=0.14$ T, 
 $\alpha_A=0.06$ and $\alpha_B=-0.04$.
The real and imaginary parts of the magnon energy are shown in Fig.~\hyperref[Fig:2]{\ref*{Fig:2}(a)} and~\hyperref[Fig:2]{\ref*{Fig:2}(b)}, respectively.  We find two regions enclosed by EPs: region I near $J=0$ and region II near $J=12.2$ $\mu$eV, i.e., the exchange interaction of CrCl$_3$~\cite{macneill2019gigahertz}.  Region I corresponds to  region I shown in Figs.~\hyperref[Fig:1]{\ref*{Fig:1}(b)} and~\hyperref[Fig:1]{\ref*{Fig:1}(c)}. Region II emerges instead in correspondence with a  noncollinear ground state and, as we will show in details, its nonlinear magnetization dynamics~(\ref{eq:simple-eg-2a},\ref{eq:simple-eg-2})  display very different features from the ones observed in region I.

Figures~\hyperref[Fig:2]{\ref*{Fig:2}(c)} -~\hyperref[Fig:2]{\ref*{Fig:2}(e)} show the time evolution of the relative alignment of the macrospins $\mathcal{S}_{AB}(t)$ for, respectively, $J=9$, $12.2$, and $16$ $\mu$eV.  Similarly to region I, passing through the EPs yields a dynamical phase transition. However, 
around region II, the exchange interaction is too strong for  a FM-to-AFM switching to take place. Instead, while for $J=9.0$ $\mu$eV and $J=16.0$ $\mu$eV we observe damped dynamical phases, see Figs.~\hyperref[Fig:2]{\ref*{Fig:2}(c)} and ~\hyperref[Fig:2]{\ref*{Fig:2}(e)}, inside region II (i.e., $J=12.2$ $\mu$eV) a periodic dynamical phase emerges, as shown in Fig.~\hyperref[Fig:2]{\ref*{Fig:2}(d)}. 
Within the periodic dynamical phase, the value of $\mathcal{S}_{AB}$ ranges from $0.7$ to $-0.7$, signaling unusual large-amplitude oscillations. Our results show that, although the overall loss  is larger than the effective gain, i.e.,  $\alpha_{A}>\abs{\alpha_{B}}$, the system can still survive in a steady periodic state in a EP-enclosed region. The dynamical phase transition can be understood as a supercritical Hopf-Bifurcation~\cite{kim2012spin,strogatz2018nonlinear,mayergoyz2009nonlinear}. When crossing the EPs and entering in region II, the fixed point of the dynamical system, which corresponds to the damped magnetization dynamics, bifurcates into a stable orbital. We have verified numerically that the large-amplitude oscillations persist at long times.

\textit{Tunability.} We proceed to investigate the dependence of the periodic stable magnetization dynamics on the system's parameters. Not surprisingly, the stability of the periodic solution strongly depends on the ratio between the effective gain and loss. Setting $J=12.2$ $\mu$eV and  $\alpha_A=0.06$, in Fig.~\hyperref[Fig:3]{\ref*{Fig:3}(a-d)} we show the time evolution of $\mathbf{S}_{A}$ (upper panel) and $\mathbf{S}_{B}$ (lower panel) on the Bloch sphere decreasing the effective gain  $\abs{\alpha_B}$ from $0.055$ to $0.01$. The colors in Fig.~\hyperref[Fig:3]{\ref*{Fig:3}(a-d)} are in direct correspondence with the time intervals of the time-evolution of $\mathcal{S}_{AB}$ shown in Fig.~\hyperref[Fig:2]{\ref*{Fig:2}(c-e)}. 
For larger values of gain, e.g., $\alpha_{B}=-0.055$, the dynamics of both macrospins $\mathbf{S}_A$ and $\mathbf{S}_B$  flow to a fixed point, as shown by Fig.~\hyperref[Fig:3]{\ref*{Fig:3}(a)}. We have verified that the same scenario is realized at the $\mathcal{PT}$-symmetric point. For lower values of the  gain, the spin dynamics evolve into a steady-state oscillations, see Figs.~\hyperref[Fig:3]{\ref*{Fig:3}(b-d)}. 
Since the macrospin $\mathbf{S}_B$ is directly subjected to gain  while $\mathbf{S}_A$ experiences it indirectly via the coupling to $\mathbf{S}_A$, the amplitude of oscillations of the macrospin $\mathbf{S}_A$ is smaller than the one of $\mathbf{S}_B$. For decreasing $\alpha_B$, the amplitude of both limit cycles shrink. 
 
In an experimental setup, the effective gain $\alpha_B$ can be controlled via the injection of spin current $J_{s}$ into the bottom layer. As shown in a very recent work~\cite{duine2023non}, swapping the dynamical gain in Eq.~(\ref{eq:simple-eg-2}) with a spin-transfer torque term, i.e., $-\frac{\alpha_B}{S}\mathbf{S}_B\times \frac{\mathrm{d}\mathbf{S}_B}{\mathrm{d}t} \rightarrow J_{s} \mathbf{S}_{B} \times (\mathbf{S}_{B} \cross \hat{\mathbf{z}}) $ does not affect the emergence of an oscillatory phase in correspondence of EP crossing.

The ratio $\alpha_A/\alpha_{B}$ is determined by the spin current transport efficiency through the magnetic layers  which, to our knowledge, has not been yet thoroughly investigated in van der Waals magnets. It is worth noting that here we take CrCl$_3$ as an example; in practice, the high degree of tunability offered by synthetic AFMs might make them a more desirable platform for engineering non-Hermitian phenomena~\cite{subedi2023magnon}. To avoid spin current injection in the top layer of a synthetic AFM bilayer, one could sandwich a good spin sink, e.g., Pt thin film~\cite{liu2019observation,heinrich2003dynamic}, between the two magnetic layers. In this case, the strength of the (RKKY) interlayer coupling can be controlled by tuning the Pt layer thickness~\cite{liu2019observation}. Synthetic AFM based on permalloy magnetic elements display an easy-plane anisotropy consistent with our model~(\ref{eq:simple-eg-1})~\cite{subedi2023magnon}.

We find that the periodic oscillatory phase does not require fine-tuning but it can instead be accessed within a relative broad range of $\alpha_A/\alpha_{B}$ values.  As shown in Fig.~\hyperref[Fig:3]{\ref*{Fig:3}(e)}, the strength of the interlayer coupling controls the frequency $f$ of the periodic oscillations (found by changing $B_0$) of the coupled dynamics $\mathcal{S}_{AB}$. For CrCl$_3$~\cite{macneill2019gigahertz}, the interlayer coupling strength $J=12.2$ $\mu$eV yields large-amplitude oscillations with frequencies in the $1-10$ GHz range. 

Finally, we explore the dependence of the onset of region II on the strength and direction of the applied magnetic field. In Fig.~\hyperref[Fig:3]{\ref*{Fig:3}(f)}, we plot  $\mathcal{P}_{EP}\equiv \abs{\bra{\psi_1^R}\ket{\psi_2^R}}^2$, where $\psi_{1,2}^R$ are the two right eigenvectors of the non-Hermitian Hamiltonian $\mathcal{H}_{nh}$. While approaching an exceptional point, the two eigenstates coalesce, i.e., $\mathcal{P}_{EP} \rightarrow 1$. The two red regions in Fig.~\hyperref[Fig:3]{\ref*{Fig:3}(f)} appear in proximity of the EPs: the region comprised between them, which centers on white and blue, corresponds to region II, i.e., it displays periodic oscillatory coupled spin dynamics. As shown by Fig.~\hyperref[Fig:3]{\ref*{Fig:3}(f)}, accessing the region II does not require  fine-tuning: there is a broad range of values of the magnetic field's strength and polar angle $\Theta$, with $\mathbf{B}_0 \cdot \hat{\mathbf{z}} =B_0 \sin \Theta$, for which the steady-state oscillations appear.


\textit{Discussion and outlook}. In this work, we investigate the interplay between the linear and nonlinear spin dynamics in proximity of exceptional points. We show that the emergence of EPs in the linearized magnon Hamiltonian underlies a dynamical phase transition of the nonlinear spin dynamics. As an example, we consider on an easy-plane bilayer in which, while one layer experiences effective gain, the other layer keeps larger loss rate.  An analysis of the linearized long-wavelength magnetization dynamics of the bilayer shows that two regions encircled by EPs can appear as function of the interlayer coupling. One region, characterized by small values of the interlayer coupling, displays an interlayer FM-to-AFM dynamical phase transition. The second region, appearing for larger values of the AFM interlayer coupling, displays large-amplitude steady-state oscillations without fine-tuning or $\mathcal{PT}$ symmetry. We argue that this oscillatory dynamical regime might be accessed via spin injection in $\text{CrCl}_{3}$ or synthetic AFM bilayers,   opening a concrete route  for experimentally probing magnetic EPs and for engineering large-amplitude magnetic nano-oscillators. 

Our theory has the potential to shed light onto the relation between non-Hermitian singularities and dynamical phase transitions in a plethora  dissipative-driven systems whose dynamics display non-linearities, e.g., molecular spin dimers~\cite{ardavan2015engineering,ramsey2008quantum}, quantum dots~\cite{amin2009coherent,zhang2003spin,yang2013pure} and  microwave resonators~\cite{PhysRevLett.120.113901,Gunnink2022}.

\textit{Acknowledgments}. K. Deng thanks B. Li for helpful discussions. This work was supported by  the National
Science Foundation under Grant No. NSF DMR-2144086.

\bibliographystyle{apsrev4-1}

\end{document}